\documentclass[11pt,twoside]{article}
\usepackage{asp2004}
\usepackage{psfig}
\usepackage{epsf}
\newcommand{\rem}[1]{} 
\begin{document}

\title{Collisions Between Single Stars in Dense Clusters: Runaway Formation of a Massive Object}

\author{Marc Freitag}
\affil{Astronomisches Rechen-Institut, Heidelberg, Germany}
\author{M. Atakan G\"urkan and Frederic A. Rasio}
\affil{Department of Physics and Astronomy, Northwestern University, USA}

\begin{abstract}
  Using Monte Carlo codes, we follow the collisional evolution of
  clusters in a variety of scenarios. We consider the conditions under
  which a cluster of main sequence stars may undergo rapid core
  collapse due to mass segregation, thus entering a phase of runaway
  collisions, forming a very massive star (VMS, $M_\ast > 1000\,{\rm
    M}_\odot$) through repeated collisions between single stars.
  Although collisional mass loss is accounted for realistically, we
  find that a VMS forms even in proto-galactic nuclei models with a
  high velocity dispersion (many 100\,km\,s$^{-1}$). Such a VMS may be
  a progenitor for an intermediate-mass black hole ($M_\bullet\ge
  100\,{\rm M}_\odot$). In contrast, in galactic nuclei hosting a
  central massive black hole, collisions are found to be disruptive.
  The stars which are subject to collisions are progressively ground
  down by high-velocity collisions and a merger sequence appears
  impossible.
\end{abstract}

\section{Introduction}
In a stellar cluster, if the velocity dispersion is smaller than about 
300\,km\,s$^{-1}$, the collision cross section is dominated by gravitational
focusing. Under these conditions the (local) collision time, i.e., the 
average time after which all stars have experienced one collision, is given by
\begin{equation}
 t_{\rm coll} \simeq 2.1\times 10^{12}\,{\rm yr}\,\frac{10^6\,{\rm pc}^{-3}}{n} 
\frac{\sigma_v}{30\,{\rm km\,s}^{-1}} \frac{{\rm R}_\odot}{R_\ast} \frac{{\rm M}_\odot}{M_\ast},
\label{eq:t_rlx}
\end{equation}
where $n$ is the stellar density, $\sigma_v$ the 1D velocity
dispersion, and $R_\ast$ and $M_\ast$ are the typical values of the radius and
the mass of a star, respectively. 
In view of this relation, one may think that direct
collisions never play a role in cluster dynamics unless they are
mediated by binary interactions, in which case the cross section is
larger by a factor $a/R_\ast$ where $a$ is the typical binary
separation. The role of binaries in cluster dynamics is presented 
by Fregeau, McMillan and Portegies-Zwart in these proceedings. 
In contrast, here we focus on two cases in which
the collisions between single stars are likely to play an important role:
(1) fast core collapse followed by runaway collisions in young dense
clusters and (2) stellar dynamics in the vicinity of a massive black
hole (MBH) at the center of a galaxy. In case (1), if there are no
primordial binaries, the merger sequence is likely to start before the
first binary forms dynamically through 3-body interactions
\citep*{FGR04}; furthermore, primordial binaries probably foster
collisions rather than preventing them by stalling the 
core collapse. In case (2), 
interaction rate peaks in the most central parts where
densities exceed $10^7\,{\rm pc}^{-3}$. However, at this location
owing to the MBH's presence, the velocity dispersion
is also very high; as a result, most binaries
are soft and should be dynamically dissociated.

To simulate the evolution of stellar clusters containing a large
number of stars over time scales comparable to the relaxation (or
collision) time, we use two Monte Carlo (MC) codes based on the
pioneering ideas of \citet{Henon71a,Henon71b}. These codes are
described in detail in the literature 
\citep{JRPZ00,JNR01,FB01a,FB02b}. The MC method assumes that the cluster is
spherical, in dynamical equilibrium and that 2-body relaxation can be
treated in the Fokker-Planck approximation. The cluster is represented
as a set of spherical shells (``particles'') each of which may stand for a single
star or a fixed number of stars which share the same orbital and stellar
properties\footnote{When one or a few particles start playing a
  special role, as in the collisional runaway process, using a
  a single particle to represent multiple stars is clearly
  questionable.}. Typical particle numbers used in the simulations range
between $10^5$ and $10^7$. The stellar orbital motion is not followed
explicitly; instead one tracks the long-term evolution of orbits (and
stellar properties) subject to relaxation, direct stellar collisions,
stellar evolution and possibly interactions with a central MBH. 
Even though the effect of binaries may be
accounted for in the MC method \citep{FregeauEtAl03}, here we consider
only single stars. Various
prescriptions can be used for the outcome of stellar collisions,
ranging from sticky spheres to inter/extrapolation from a large
database of SPH simulation results \citep{FB04}.

\section{Runaway Formation of a Very Massive Star}

We consider the evolution of young stellar clusters with a broad
initial mass function (IMF). For all results presented here, we used
the Salpeter IMF, $dN_\ast/dM_\ast \propto M_\ast^{-2.35}$ from $0.2$
to $120\,{\rm M}_\odot$. There is no initial mass segregation.

%\subsection{Fast core collapse}

\begin{figure}[!ht]
%black-white version %\plotone{freitag_fig1.eps}% Mcl_Rh_runaway_plane.eps --> freitag_fig1.eps
\plotone{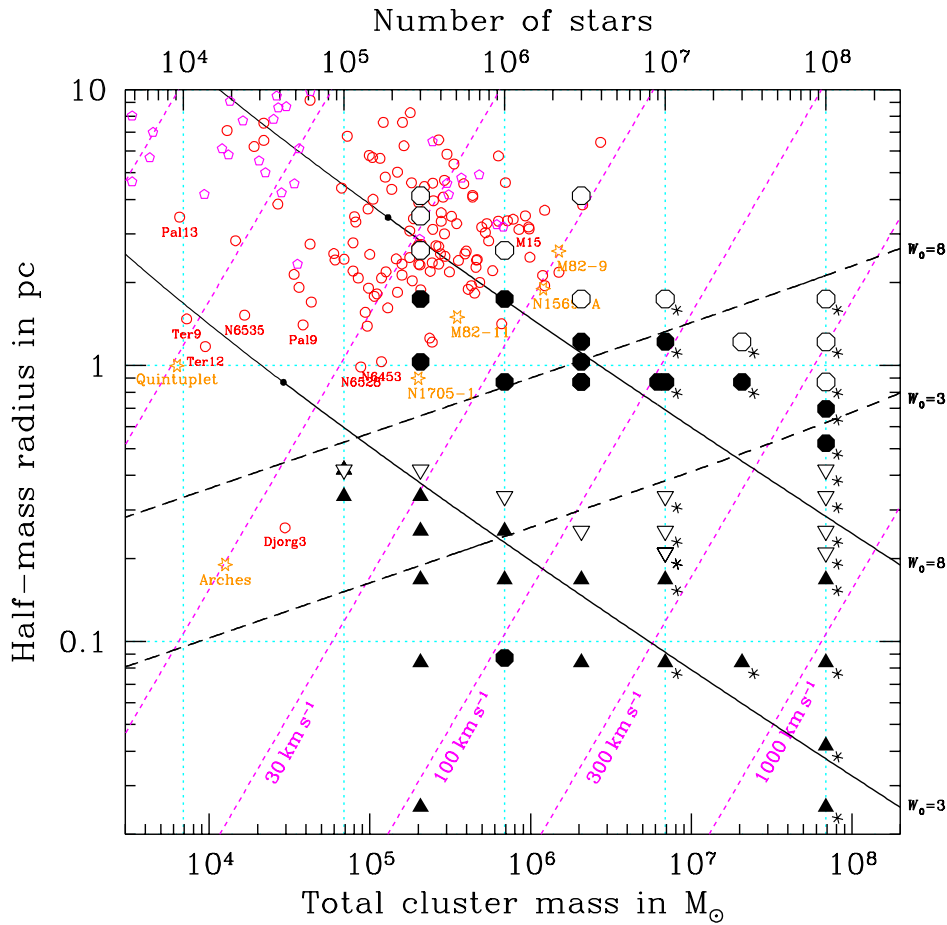}
\caption{
  Conditions for quick core collapse and collisional runaway. Solid
  lines show the condition that the time for core collapse by relaxation is
  3\,Myr for King models with $W_0=3$ and $W_0=8$ and our standard
  IMF. Below this line, core collapse time is shorter. Dots on these
  lines indicate models that have initially $10^4$ stars in their 
  core, i.e. $\sim 4$ stars more massive than $50\,{\rm M}_\odot$!
  Long-dashed lines indicate where the collision time for a 120$\,{\rm
    M}_\odot$ star is 3\,Myr. Below this line, one may expect
  collisional effects to be important relative to relaxation.  The
  approximate central 1D velocity dispersion is also indicated.  
  Small symbols indicate the estimated conditions of a variety of
  observed clusters (globular or young); large symbols indicate the
  initial conditions of our MC simulations (triangles and round symbols
  for $W_0=3$ and 8, respectively). Open symbols are for runs which missed the
  runaway phase, and the filled symbols for those that experienced it.
  Asterisks denote simulations where the number of particles are smaller than
  the number of stars.}
\label{fig:MRplane}
\end{figure}

We first studied the relaxation-driven core collapse to determine the
conditions for it to occur while the most massive stars are still on
the main sequence (MS), i.e. within $t_{\rm MS}(50-120\,{\rm M}_\odot)
\approx 3$\,Myr \citep*{GFR04}.  This condition arises because the
mass loss during the giant phase and the subsequent supernova
explosions would stop the core collapse and cause the cluster to
expand. A key finding of our work is that, for a given cluster
structure and a ``reasonable'' IMF (neither too flat or too steep),
the core collapse time, $t_{\rm cc}$, is simply proportional to the
initial central relaxation time, where the proportionality constant
depends on the ratio of the maximum to the average stellar mass,
$t_{\rm cc} \simeq \alpha(\mu)\,t_{\rm rc}(0)$ with $\mu=M_{\rm
  max}/\langle M\rangle$.  Furthermore, for $\mu>50$, a regime reached
by all realistic IMF, the dependency flattens to a constant value:
$t_{\rm cc} \simeq 0.15\,t_{\rm rc}(0)$, independently of the cluster
structure. With this we can predict that in any cluster with $t_{\rm
  rc}(0)\le 20$\,Myr, massive stars will form a dense collapsing core
before they turn into remnants and should start colliding with each
other. This condition for quick core collapse is plotted in
Fig.~\ref{fig:MRplane}. We see that, if clusters are born with
relatively high concentration, a significant fraction of them lie in
the proper region of parameter space.

%\citep{GFR04}

%\subsection{runaway sequences}

More recently we have explicitly considered the effects of collisions
in a series of MC simulations \citep{FGR04}.  As shown on
Fig.~\ref{fig:MRplane}, quick core collapse, followed by the runaway
growth of a VMS through repeated mergers, happened in all cases for
which we predicted $t_{\rm cc}<\,3$\,Myr, except when the number of
particles is too low to have a critical number of massive stars
initially present in the core. For clusters with large masses and small
sizes, collisions themselves shorten the effective $t_{\rm cc}$ and we
obtained runaway growth for larger clusters than predicted from the
simple relaxational core collapse picture.

\begin{figure}[!ht]
%black-white version %\plotone{freitag_fig2.eps}% merger_trees.eps --> freitag_fig2.eps
\plotone{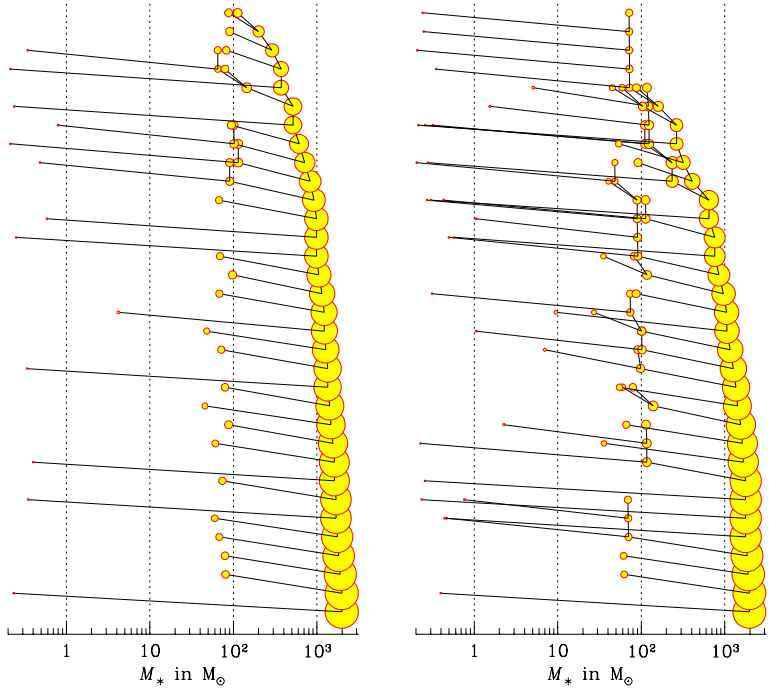}
\caption{Merger trees for the simulation of clusters with $W_0=3$,
  $N_{\ast}=N_{\rm part}=3 \times 10^5$. Left: $R_{\rm h}=0.48$\,pc.
  Right: $R_{\rm h}=0.036$\,pc. We follow the growth of the runaway
  star to $\sim 2000\,{\rm M}_\odot$. In case (a), the cluster is not
  initially collisional and most collisions occurs in deep collapse
  and feature stars of mass $70-120\,{\rm M}_\odot$ that have
  segregated to the center. In case (b), the cluster is initially
  collisional and most stars contributing to VMS growth have
  experienced earlier collisions.}
\label{fig:trees}
\end{figure}

\begin{figure}[!ht]
%black-white version %\plotone{freitag_fig3.eps}% coll_histories.eps --> freitag_fig3.eps
\plotone{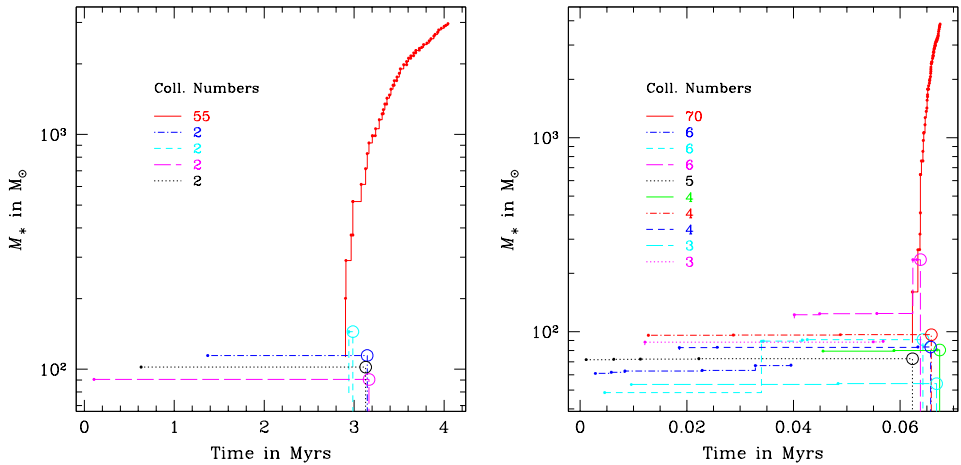}
 \caption{ 
   Collisions histories for the same simulations as in
   Fig.~\ref{fig:trees}. Circles indicate that the particle has merged
   with the runaway object which grows very quickly at the moment of
   core collapse. For the smaller cluster, the collapse is so fast that
   it should be considered in the the frame of the formation process
   of the cluster.}
\label{fig:masses}
\end{figure}

\begin{figure}[!ht]
%black-white version %\plotone{freitag_fig4.eps}% tscales_runaway.eps --> freitag_fig4.eps
\plotone{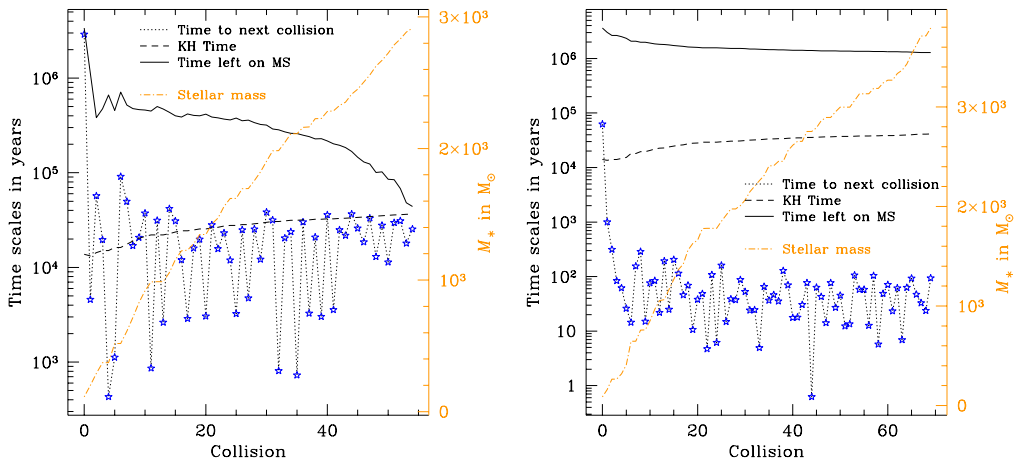}
 \caption{ 
   Time scales and mass evolution during the growth of the runaway
   star for the simulations of Fig.~\ref{fig:trees}. We plot the time
   between successive collisions, an estimate of the (MS)
   Kelvin-Helmholtz time scale $T_{\rm KH}$, the time left until
   exhaustion of the central hydrogen (left scale) and the mass of the
   star (right scale). $T_{\rm KH}$ is not considered during the
   simulations.}
\label{fig:tscales}
\end{figure}

Figures~\ref{fig:trees}, \ref{fig:masses} and \ref{fig:tscales}
present a detailed look at the runaway sequences obtained in two
simulations. In these cases we used the sticky sphere approximation
which we checked to be fully justified when $\sigma_v\le
300$\,km\,s$^{-1}$. We note that, in most situations, a very important
fraction of the mass accumulated in the VMS comes from the upper-most
part of the IMF, $M_{\ast} \sim 100\,{\rm M}_\odot$, due to strong
mass segregation.  Another important finding is that, generally, the
interval between mergers is much shorter than the thermal time scale
of the VMS.  Consequently, its structure is probably more extended and
diffuse than a MS star, contrary to our assumption. Average growth
rates are typically higher than $10^{-3}\,{\rm M}_\odot{\rm yr}^{-1}$.

%\subsection{Open questions}

Our results raise a number of open questions. One important planned
extension of this work is to establish the role of primordial
binaries. Most other uncertainties are connected with the final mass
of the VMS and its fate. The vast majority of our simulations were
artificially terminated when the VMS reached some high mass (typically
$2000\,{\rm M}_\odot$) while the average time between mergers was
still much shorter than the remaining MS lifetime. We think that,
although they include collisional mass loss, our models lack the
physics limiting the growth of the VMS. Preliminary models assuming a
fixed central VMS but accounting for depletion of ``loss-cone'' orbits
show a saturation at a few
$1000\,{\rm M}_\odot$. %show a saturation of $M_{\rm VMS}$. 
Another limiting mechanism may be that the VMS cannot radiate
collision energy and swells until it becomes ``transparent'' to
impacting stars. Whether the VMS will spawn an intermediate-mass BH is
also uncertain because, once collisional growth levels off, strong
stellar winds or pulsational instabilities may strongly reduce the VMS
mass, unless the metallicity is very low. Finally we note that since
the process occurs on a Myr time scale, it should be considered in the
context of cluster formation, with account for the residual gas and
pre-MS stars.

\section{Collisions in Galactic Nuclei}

\begin{figure}[!ht]
%black-white version %\plotone{freitag_fig5.eps}% CollHistMass_PC_1187.prn.eps --> freitag_fig5.eps
\plotone{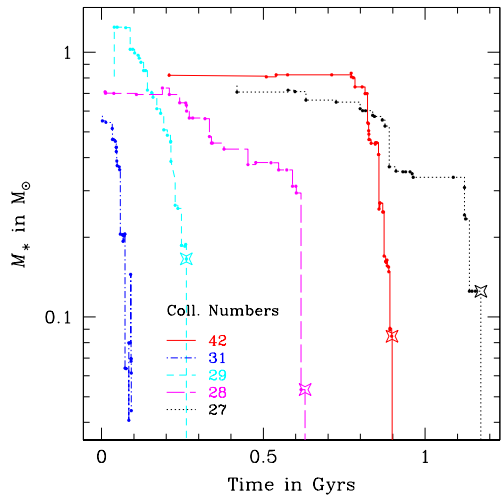}
 \caption{ 
   Collision sequences in the same simple model of the SgrA$^\ast$
   cluster as used by \citet{Freitag03}.
   %The top panel indicates at which distance from
   %the center each collision occurred. The bottom panel shows the mass
   %evolution for the same particles.
}
\label{fig:galnuc}
\end{figure}

The situation is completely different in galactic nuclei harboring a
central MBH. Inside the radius of influence of the MBH, the stellar
density probably follows a power-law cusp $n\propto R^{-\gamma}$ with
$\gamma \simeq 1.5-2$ and the velocity dispersion raises like
$\sigma_v\propto R^{-1/2}$. Hence, $t_{\rm coll} \propto
R^{-(\gamma\pm 1/2)}$ and collisions should be effective very close to
the MBH. But there the relative velocities are so high that partial or
complete disruptions are more likely than mergers. Even a short
sequence of mergers was never found in older MC galactic nucleus simulations
such as shown in Fig.~\ref{fig:galnuc}.

\acknowledgments{MF acknowledges support by the scheme SFB-439/A5 of the
  German Science Foundation (DFG). The work of AG and FR is supported
  by NASA ATP Grant NAG5-12044 and NSF Grant AST-0206276.}

%\bibliographystyle{apj}
%\bibliography{aamnem99,biblio}
%\rem{

%}

\end{document}